# Inverse design of compact optical cloaks and experimental demonstration at microwave frequencies

**Mediha Tutgun[1], Emre Bor[1,2], Mirbek Turduev[2] and Hamza Kurt[1]**

[1]Department of Electrical and Electronics Engineering, TOBB University of Economics and Technology, Ankara 06560, Turkey

[2]Department of Electrical and Electronics Engineering, TED University, Ankara 06420, Turkey

*Corresponding author: <mtutgun@etu.edu.tr>

**Inverse design in photonics has gathered increasing attention as a powerful approach that goes beyond the intuition-based designs. In this Letter, we present the inverse design and experimental demonstration of compact optical cloaks at microwave frequencies is conducted. Two different configurations of rectangular and circular cloaks are numerically designed to reduce the scatterings of incident light interacting with a perfect electrical conductor object. The designed cloaking structures consist of dielectric polylactide material with a low refractive index and they are fabricated by 3D printing approach. The experimental measurements are in good agreement with the numerical calculations. The designed region covering ($4\lambda$ x $4\lambda$) area enables hiding circular object of $\lambda$ diameter where $\lambda$ denotes the wavelength of incident light. The proposed approach may enable the concealment of different objects possessing various size and shapes.**

Optical cloaking is the phenomenon that provides the invisibility effect of objects at selected operating wavelengths [1, 2]. The revelation of optical cloaking phenomenon can be considered as one of the most alluring and intriguing optical applications. In the pioneer works, transformation optics and conformal mapping are applied to obtain invisibility [3,4]. In the last decade, various studies dealing with optical cloaking have been reported. For example, quadruple Luneburg lens system is presented to achieve a hidden region between lenses [5]. Also, a carpet cloaking effect is achieved by employing semi-Dirac cone dispersion of a photonic crystal structure [6]. Furthermore, optimization algorithms such as topology optimization and genetic algorithm are applied to achieve directional invisibility [7, 8]. Additionally, metasurfaces and zero-index materials are combined to design hybrid invisibility cloaks [9]. In recent years, additive manufacturing methods are utilized to design optical cloaks at microwave frequency regime [5, 8, 10, 11].

Recently, inverse design approaches enabled the design of efficient optical devices with small footprints [12, 13]. For this purpose, several inverse design approaches including machine learning are proposed for optical designs [14-16]. The presented studies of inversely designed optical structures have gathered great interest and are promising for the future of optical designs [17,18].

In this study, we propose the design of compact optical cloaking devices by applying an objective-first inverse-design (OFID) strategy. In the literature, the OFID approach has been utilized to define complex photonic structures with high degrees of freedom [12, 13]. In the proposed design strategy, target parameters defined in the cost function as a boundary-value problem, and the permittivity ($\varepsilon$) distribution of the structure is obtained iteratively.

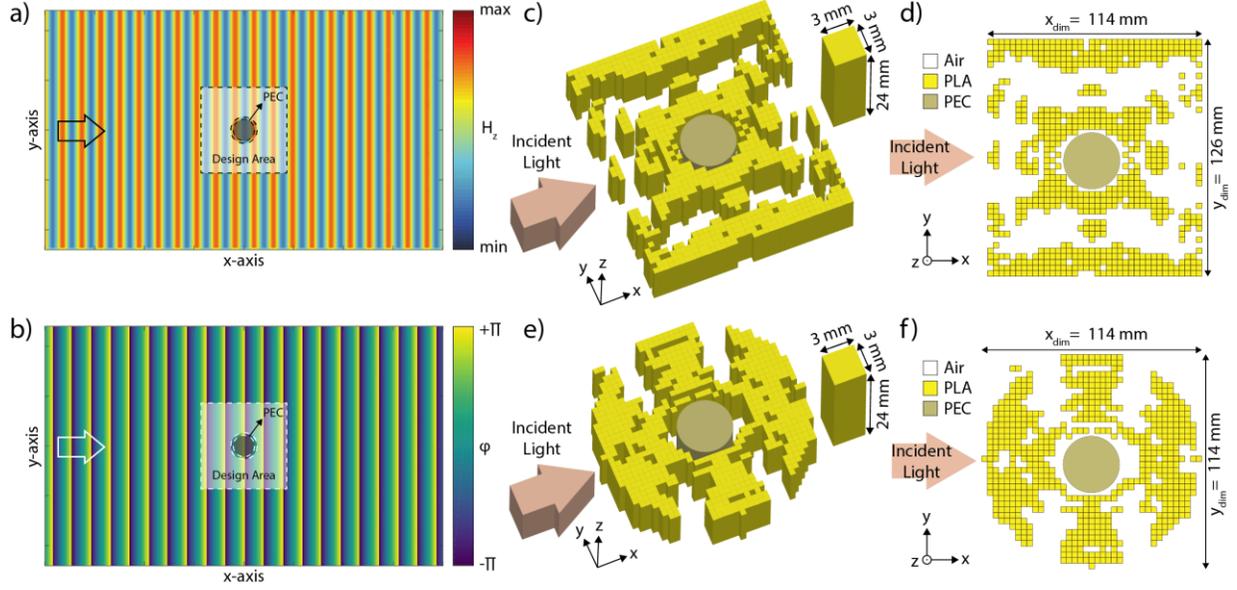

**Fig. 1.** Schematic representation of design objectives for (a) field and (b) phase distributions. (c) Perspective and (d) top views of rectangular cloak with its structural parameters. (e) Perspective and (f) top views of circular cloak with its structural parameters.

The problem of invisibility is treated as an inverse problem in contrast to the conventional forward design approaches based on prior knowledge. In this strategy, the desired output electromagnetic field is defined, and the algorithm iteratively searches for the best structure that provides the desired output. In other words, the inverse design method modulates the permittivity distribution in the optimization region, i.e., design area, to obtain desired light-matter interaction at the defined operating wavelength. It should be noted that the optimization region is divided into smaller rectangular areas namely pixels.

Throughout the optimization process, the permittivity value of polylactide (PLA) material is fixed to $\varepsilon_{PLA} = 2.4025$ whereas the permittivity of air background is equal to $\varepsilon_{PLA} = 1.0$. The upper and lower limits of permittivity values for pixels are determined as $\varepsilon_{air} \leq \varepsilon \leq \varepsilon_{PLA}$ to design optical cloaks.

Due to the operation of OFID algorithm, the resulted solution has continuous permittivity distribution. Later, Sigmund-Strength discretization method is applied to convert continuous permittivity distribution to discrete permittivity distribution without sacrificing the cloaking performance [19]. As a result, pixels of optical cloaks can have the permittivity values of PLA or air which enables the fabrication of structures. Here, the light-matter interaction of the final structure the is analyzed by using the finite-difference frequency-domain (FDFD) method [12]. At the final stage, the optimized structures are investigated in the time domain via three-dimensional (3D) finite-difference time-domain (FDTD) method [20].

In this study, we designed two different optical cloaking structures with compact sizes which have circular and rectangular shapes. The designed all-dielectric structures consist of PLA material which has a low refractive index of $n_{PLA}$=1.55 [8]. Here, we defined a design area to surround a perfectly electrical conductor (PEC) cylindrical object. In order to design an optical cloak, plane wave nature of incident light should be conserved behind the cloaking structure in terms of magnetic field and phase distributions which are schematically represented in Fig. 1(a) and 1(b), respectively. For this purpose, we set targets for optimization process as straight cross-sectional profiles of magnetic field and phase distributions at the output design area. In these figures, the dark shaded circular region represents the PEC cylindrical object whereas the light shaded region is defined as design area where the inverse design approach determines the permittivity values of pixels.

The 3D perspective and top views of designed rectangular cloak are represented in Fig. 1(c) and 1(d) with its structural parameters. The designed rectangular cloak has a dimension of $x_{dim}$=114 mm in the longitudinal direction and $y_{dim}$=126 mm in the lateral direction. Additionally, in Fig.1(e) and 1(f), perspective and top views of circular cloak are given respectively. The designed circular cloak has dimension $x_{dim}$=114 mm and $y_{dim}$=114 mm, respectively, in longitudinal and lateral directions. The corresponding structures consist of dielectric rectangular rods, i.e., pixels, with dimensions of 3×3 mm$^2$ and a thickness of 24 mm.

The distributions of perpendicular magnetic field component ($H_z$) are calculated via 3D FDTD method in order to compare the cases of PEC object and cloaked PEC object by rectangular design and circle design in Fig 2. Firstly, for the case of rectangular cloak, the magnetic field and phase distributions of PEC object and cloaked PEC at 10.74 GHz are given in Fig 2(a) and 2(b), respectively. Here, for the cases of both cloaking structures, the cylindrical PEC object has a diameter of 30 mm and a thickness of 24 mm. As can be seen from these figures, the undesired scatterings are suppressed, and the magnetic field distribution at the backplane of the structure is corrected which resembles the incident plane wave. Also, the calculated cross-sectional amplitude and phase profiles are shown in Fig. 2(c) and 2(d), respectively. It can be seen that the fluctuations in the case of PEC only simulation are suppressed when the PEC object is surrounded by the rectangular cloak.

In addition, we performed the similar analyzes for the cases of bare PEC and cloaked PEC by circular structure at 10.16 GHz. The corresponding magnetic field and phase distributions are represented in Fig. 2(e) and 2(f). Moreover, the cross-sectional profiles of magnetic field and phase distributions at input and output are plotted in Fig. 2(g) and 2(h), respectively, for the cases of PEC object and concealed PEC object. It can be concluded that the circular cloak is able to flatten the fluctuations in amplitude and phase profiles. Furthermore, the transmission efficiencies are calculated as 85% and 87% for rectangular cloak and circular cloak, respectively. It can be concluded from Fig. 2 that the proposed inverse design approach can provide a favorable solution for cloaking purposes. The designed structures have compact sizes comparing to the diameter of PEC object.

For experimental verification of numerical calculations at microwave frequency regime, the inversely designed cloaks are fabricated by a 3D printing technique which utilizes PLA material. PLA is an extensively used thermoplastic material for rapid prototyping. A brass cylindrical object is used for the substitution of PEC object in experimental demonstration. Brass is the compound of copper and zinc materials, which has scattering properties at microwave frequencies between 8–12 GHz. Throughout the experimental process, the Agilent E5071C ENA vector network analyzer is used to generate and measure electromagnetic waves at frequencies of interest.

An aperture antenna (horn antenna) is placed at a distance to ensure that the incident waves have planar wavefronts at the front surface of cloaking devices. Due to physical restrictions of the experimental setup, we were able to scan the area with sizes of 20×20 mm² behind the cloaking devices by using a monopole antenna. In Fig. 3(a), the schematic representation of experimental setup for the case of rectangular cloak is given. The photographic view of fabricated rectangular cloak and brass cylindrical object is shown in Fig. 3(b). Firstly, we positioned brass object only and measured the magnetic field and phase distributions behind it at frequency around 10.34 GHz. Later, we placed the rectangular cloak to surround the brass object and realized same measurements. The corresponding magnetic field and phase distributions are presented in Fig. 3(c) and 3(d), respectively. In Fig. 3(e), the cross-sectional amplitude and phase profiles at the output are plotted for the cases of object only and concealed object by rectangular cloak.

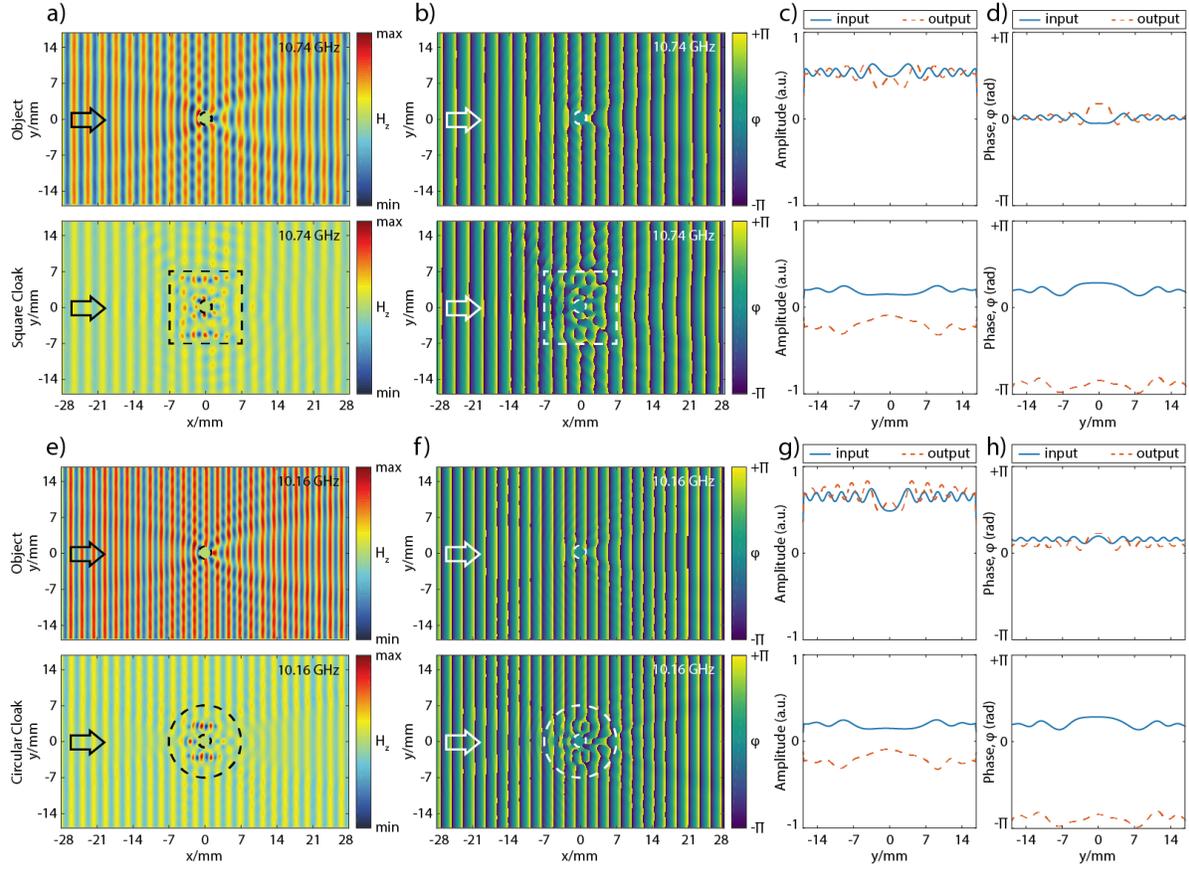

**Fig. 2.** Calculated (a) magnetic field and (b) phase distributions, and cross-sectional (c) amplitude and (d) phase profiles for the cases of only PEC object and PEC with rectangular cloak at 10.74 GHz. Calculated (e) magnetic field and (f) phase distributions, and cross-sectional (g) amplitude and (h) phase profiles for the cases of only PEC object and PEC with circle cloak at 10.16 GHz. The arrows indicate the propagation direction of incident plane waves. The outer boundaries of PEC objects and cloaks are denoted by dashed lines.

In addition, we experimentally demonstrated the cloaking ability of circular cloak at frequencies around 9.76 GHz by using the experimental setup represented in Fig. 3(f). The photographic view of circular cloak is given in Fig. 3(g). Here, we measured the magnetic fields behind the brass object and repeated the same measurement when the brass object is placed in the center cavity of circular cloak. The corresponding magnetic field distributions are presented in Fig. 3(h). Also, the corresponding phase distributions are measured and depicted in Fig. 3(i). Later, in Fig. 3(j), we plotted the cross-sectional amplitude and phase profiles for both cases.

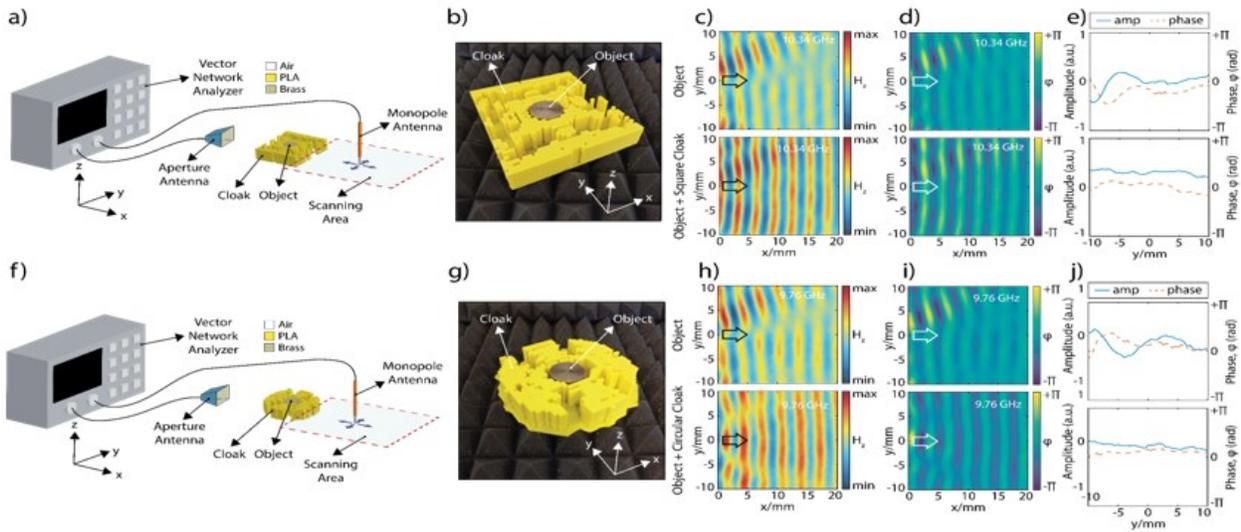

Fig. 3. (a) Schematic representation of experimental setup and (b) photographic view of rectangular cloak. Measured (c) magnetic and (d) phase distributions, and (e) corresponding cross-sectional profiles for the cases of only brass object and object with rectangular cloak at 10.34 GHz. (f) Schematic representation of experimental setup and (g) photographic view of circular cloak. Measured (h) magnetic and (i) phase distributions, and (j) corresponding cross-sectional profiles for the cases of only brass object and object with circular cloak at 9.76 GHz. The arrows indicate the propagation direction.

It can be concluded from the experimental measurements that both rectangular and circular cloaks are able to conceal brass cylindrical object by suppressing the scatterings of light at different microwave frequencies. Additionally, the presented cross-sectional profiles prove the cloaking ability of proposed structures. Here, the field deteriorated by scatterer object is reformed by inversely designed cloaks.

The inversely designed cloaking structures exhibit an invisibility effect for scatterer objects by providing the same field and phase properties at the output of the structure with the ones of the incident light. The physical mechanism of achieved invisibility can be related to the intelligent distribution of permittivity values. In conformal mapping, continuous variation of permittivity values of pixels is necessary which makes the designed cloaks unrealistic for the fabrication process [3,4]. Therefore, we have applied binarization to obtained permittivity distribution while keeping the deterioration of cloaking ability at a reasonable level. As a result, we have obtained fabricable cloaking structures which have pixels with discrete permittivity values that can be air or dielectric material. Lastly, although the algorithm performs pixel by pixel optimization, the resulting design has a certain order. This structural arrangement is important in order to transmit the incident wave with less loss around the object and without any phase front distortion.

In this study, we propose the inverse design of compact and efficient optical cloaks and experimentally demonstrated their cloaking ability at the microwave frequency regime. A recently emerging optimization method called OFID is used to design the cloaking structures. The numerical calculations are carried out by employing the 3D FDTD method. Later, the designed cloaks are fabricated by additive manufacturing such that the structures are compact and consist of dielectric PLA material with low refractive index. Similar cloaking performance is obtained for both rectangular and circular designs. We believe that the introduced integration of inverse design approach and 3D printing fabrication technique can pave the way for designing other novel optical elements.

**Funding.** Türkiye Bilimsel ve Teknolojik Araştırma Kurumu (TÜBITAK) (116F200)